\documentclass[12pt]{JHEP3}

\usepackage{epsfig}
\epsfclipon
\usepackage{epsfig,bm}
\usepackage{amssymb,amsmath}
\def\IP{\relax{\rm I\kern-.18em P}}
\def\beq{\begin{equation}}
\def\eeq{\end{equation}}
\def\beqa{\begin{eqnarray}}
\def\eeqa{\end{eqnarray}}
\title{Inflating in a Better Racetrack}

\author{J.J.~ Blanco-Pillado,$^1$ C.P.~Burgess,$^{2,3}$
J.M.~Cline,$^4$ C.~Escoda,$^5$ M.~G\'omez-Reino,$^6$
R.~Kallosh,$^7$ A.~Linde$^7$ and F.~Quevedo$^5$
\\
$^1$ Center for Cosmology and Particle Physics,
New York University, New York, \\ \qquad NY 10003, USA\\
${}^2$ Department of Physics and Astronomy, McMaster University,\\
\qquad 1280 Main Street West, Hamilton, Ontario , L8S 4M1, Canada\\
${}^3$ Perimeter Institute, 31 Caroline Street North,
Waterloo, Ontario, \\ \qquad N2L 2Y5, Canada\\
$^4$Physics Department, McGill University, 3600 University Street,
Montr{\'e}al, \\ \qquad Qu{\'e}bec, H3A 2T8, Canada
\\
$^5$ DAMTP, Centre for Mathematical Sciences,
University of Cambridge,\\ \qquad Cambridge CB3 0WA, UK\\
$^6$ Institut de Physique, Universit\'e de Neuchatel,
Rue. A.-L. Breguet 1, CH-2000,\\ \qquad Neuchatel, Switzerland\\
$^7$ Department of Physics, Stanford University, Stanford, CA
94305--4060, USA\\
}

\received{\today}

\preprint{DAMTP-2006-20, SU-ITP-06-07\\  hep-th/0603129}

\abstract{We present a new version of our  racetrack inflation
scenario which, unlike our original proposal, is based on an
explicit compactification of type IIB string theory: the
Calabi-Yau manifold $\IP^4_{[1,1,1,6,9]}$. The axion-dilaton and
all complex structure moduli are stabilized by fluxes. The
remaining 2 K\"ahler moduli are stabilized by  a nonperturbative
superpotential, which has been explicitly computed. For this model
we identify situations for which a linear combination of the
axionic parts of the two K\"ahler moduli acts as an inflaton. As
in our previous scenario, inflation begins at a saddle point of
the scalar potential and proceeds as an eternal topological
inflation.  For a certain range of inflationary parameters, we
obtain the COBE-normalized spectrum of metric perturbations  and
an inflationary scale of $M=3\times 10^{14}$ GeV. We discuss
possible changes of parameters of our model and argue  that
anthropic considerations favor those parameters that lead to a
nearly flat spectrum of inflationary perturbations, which in our
case is characterized by the spectral index  $n_s = 0.95$.}


\begin{document}

\section{Introduction}

The past two years have seen continued progress in identifying how
cosmological inflation can arise from within string theory.
Several interesting mechanisms have been examined so far,
including brane/antibrane inflation
\cite{dvalitye}--\cite{cs}, D3/D7 brane inflation
\cite{d3d7}, DBI inflation \cite{dbi}, tachyon inflation
\cite{tachyon} and inflation driven by various kinds of
closed-string moduli \cite{bmqrz}--\cite{conlon}
or more stringy degrees of freedom \cite{shamit}. There is
continued interest in identifying where inflation can arise within
the enormous landscape of string configurations.

The Racetrack Inflation scenario of 
Ref.~\cite{racetrack} gives a particularly simple scenario, based
on a Calabi-Yau compactification of type IIB string theory having
only a single K\"ahler modulus once fluxes are used to fix the
complex-structure moduli. It was argued that the one remaining
K\"ahler modulus could be the inflaton provided the
nonperturbative superpotential for this modulus has a
double-exponential form, known in the literature as a racetrack
model \cite{krasnikov}. For appropriate choices of the parameters of this
superpotential this model can give rise to standard hill-top,
topological slow-roll inflation. This is in contrast with the
simplest single-exponential case originally discussed in the
KKLT model \cite{kklt}, which does not similarly give rise to inflation.

Even though very simple and predictive, a drawback with this
scenario is the absence of an explicit string construction which
produces the required racetrack superpotential. More generally, it
remains a challenge to derive inflation within an explicit string
configuration for which  all of the features of the complete
potential which stabilize the moduli are explicitly calculable.

In this paper we identify how inflation can be obtained using
the low-energy field theory which is known to describe the two
K\"ahler moduli of the $\IP^4_{[1,1,1,6,9]}$ model, for which Denef,
Douglas and Florea were able to provide explicit calculations of
the nonperturbative superpotential \cite{ddf}. The title of
their paper --- ``Building a Better Racetrack'' --- explains
the title of ours.
We should note, however, that although the superpotential they find 
is a sum of two exponentials, it differs from the usual racetrack 
superpotential because each exponential depends only on a different 
K\"ahler modulus. We find that the resulting scalar potential can 
nevertheless produce inflation in a racetrack-like way: through the 
slow-roll of a linear combination of the K\"ahler moduli axions
 away from a local saddle point towards a nearby
local minimum.

We describe these results in the following way. Section \ref{single} starts
with a brief summary of the original racetrack-inflation mechanism
for a single modulus,  followed in Section \ref{example} by a
description of its generalization to the two-modulus case of
interest for the $\IP^4_{[1,1,1,6,9]}$ type IIB vacuum. In Section
\ref{roll} we describe a particular choice of parameters which lead to
eternal topological inflation in this model. In Section \ref{spectrum}
we describe the derivation of the amplitude of the spectrum of scalar
metric perturbations in our model. In Section \ref{ns} we discuss a
relation between 
our parameters, the flatness of the spectrum, and the total duration
of inflation, 
and argue that anthropic considerations show some preferences for the
parameters 
leading to the flat spectrum. Our
conclusions are then discussed in Section \ref{discussion}.

\section{Single-Modulus Racetrack Inflation}\label{single}

In this section we briefly review the original proposal for
racetrack inflation within the KKLT scenario \cite{racetrack}. This
scenario is based on those vacua of type IIB string theory which
are obtained by compactifying to 4 dimensions in an $N=1$
supersymmetric way on an orientifolded Calabi-Yau manifold in the
presence of three-form RR and NS fluxes and D7 branes. The
presence of the fluxes can fix the values of the complex dilaton
field and of the various complex-structure moduli of the
underlying Calabi-Yau space \cite{gkp,sethi1}, leading to a
low-energy 4D supergravity describing the remaining K\"ahler
moduli which has the no-scale form \cite{noscale}, for which the
remaining moduli correspond to exactly flat directions of the
scalar potential.

\subsection{Vacuum Solutions}

Explicitly, the potential for these moduli is given by the
standard $N=1$ F-term potential of $N=1$ supergravity, which in
Planck units reads
\beq
    V_{F}~=~e^{K}\left( K^{i\bar\jmath} D_{i}W \overline {D_j W}
    - 3|W|^2 \right)\,,
\label{treepot} \eeq
with $i,j$ running over the various complex chiral fields,
$\varphi^i$. $K^{i\bar\jmath}$ denotes the inverse of the matrix
$K_{i\bar\jmath} = \partial_i \partial_{\bar\jmath} K$, and the
K\"ahler covariant derivative is $D_i W = \partial_i W
+(\partial_iK) W$, where $K$ is the system's K\"ahler potential
and $W$ its superpotential. For the low-energy moduli obtained
from a flux compactification typically $W = W_0$ depends only on
the complex-structure moduli, and is therefore a constant
so far as the
K\"ahler moduli are concerned. The K\"ahler function for the
K\"ahler moduli also satisfies the no-scale identity
\beq
    K^{i\bar\jmath} K_{i}  {K_j} =  3 \,,
\eeq
implying that $V_F = 0$; hence these moduli are not fixed by the
fluxes themselves.

In order to fix all moduli KKLT imagined starting with a Calabi-Yau
space having just a single K\"ahler modulus, the volume
modulus $T$. For this the K\"ahler potential $K$, obtained
neglecting possible corrections in powers of $\alpha'$ and the
string coupling\footnote{Perturbative corrections to $K$ have been
considered recently with interesting changes to the KKLT
\cite{bbcq} and racetrack \cite{racetrackalpha} scenarios. We
first neglect them here, although their inclusion is
straightforward.} has been computed to have the no-scale form
\cite{truncation}
\beq
    K=-3\log (T+T^*)\,.
\label{kpdef} \eeq
The flat direction in the $T$ direction is lifted once $W$
acquires a $T$-dependence, such as can be generated by
D3-instantons, or gaugino condensation if a suitable gauge sector
exists on one of the
D7 branes.

Writing the field $T$ in terms of its real and imaginary parts,
\beq
    T\, \equiv X+i Y\,,
\eeq
and using (\ref{treepot}) and (\ref{kpdef}), the supersymmetric
part of the scalar potential turns out to be
\beq
    V_{F}  =  \frac1{8 X^3} \left\{ \frac13\left|2X W^\prime  - 3
W\right|^2 - 3|W|^2
    \right \}\,,
\label{spot} \eeq
where $^\prime$ denotes differentiation with respect to $T$. The
supersymmetric minima of this potential are given by the solutions
of
\beq
    2X W^\prime - 3 W = 0\,,
\label{msloc} \eeq
for which it is clear that the value of the potential at the
minimum is generically negative, leading to vacua with 4D anti-de Sitter
geometry.

KKLT obtain metastable vacua having de Sitter geometry in four
dimensions by raising the potential at this minimum to 
positive values by introducing anti-D3 branes.
The presence of these branes does not introduce extra
translational moduli since their positions are fixed by the fluxes
\cite{kpv}, and so their low-energy effect is just a contribution
to the energy density of the system. Provided the antibrane
tension is small --- such as if they reside at the tip of a
strongly-warped throat --- it represents a small perturbation to
the total energy density, and the 4D scalar potential which
results is a sum of two parts
\beq
    V= V_F + \delta V \,.
\label{PotentialAsSum} \eeq
Here $\delta V$ represents the small, explicitly
nonsupersymmetric, part of the potential induced by the tension
of the anti-D3 branes.

Alternatively, this same lifting could be obtained by turning on
magnetic fluxes on the D7 branes, and using the resulting
Fayet-Iliopoulos D-term potential to raise the value of the
potential at its minimum \cite{bkq}, again leading to a result of
the form of eq.~(\ref{PotentialAsSum}), with $\delta V$ being the
D-term part of the potential induced by the magnetic field fluxes.
In either case the form of $\delta V$ is positive definite and
depends on an inverse power of the overall volume of the
Calabi-Yau space,
\beq
    \delta V = \frac{E}{X^\alpha} \,.
\eeq
for constants $E,\alpha > 0$. The coefficient $E$ is proportional
to the antibrane tension, $T_3$, gravitationally redshifted by the
local value of the warp factor. The exponent is $\alpha = 2$ if
$\delta V$ arises from anti-D3 branes sitting at the end of a
Calabi-Yau throat, or if it arises from magnetic-field fluxes on
D7 branes wrapped on cycles at the tip of such a throat. Otherwise
$\alpha = 3$, corresponding to anti-D3 branes (or magnetic flux on
D7 branes) situated in an unwarped region. For anti-D3 branes the
warped region is energetically preferred and so in what follows we
take $\alpha=2$.

Further progress requires specifying the $T$-dependence of the
superpotential, for which KKLT take the simplest form:
\beq \label{KKLTW}
    W = W_0 + A \, e^{-aT} \,.
\eeq
This choice allows minima for $V_F$ at large $X$
--- as is required by the supergravity approximation to string
theory --- provided the fluxes are arranged to ensure that $W_0$
is sufficiently small. (In the limit $W_0 \to 0$ the minima disappear,
 leading instead to a runaway potential.) The analysis
also goes through for more complicated possibilities for $W$,
however, such as if $W$ were given by modular functions as would
be expected for $N=2^*$ models \cite{egq}, or if it involved the
sum of two exponentials \cite{racetrack}
\beq
    W = W_0 + A \, e^{-aT} + B \, e^{-bT} \,.
\label{RacetrackW}
\eeq

This last superpotential includes the original KKLT scenario if
$AB=0$ and has the property that it can naturally provide minima
for $V_F$ which lie in the large-field region. (When $W_0 = 0$ the
superpotential of eq.~(\ref{RacetrackW}) reduces to the standard
racetrack form, which was studied in the past as a superpotential
which could stabilize the dilaton field at weak coupling for
heterotic string vacua \cite{krasnikov}.) Such a superpotential
arises when gauginos condense for a supersymmetric gauge theory
(with no charged matter) involving a product gauge group. For
instance, the gauge group $SU(N)\times SU(M)$ leads to a sum of
exponentials with both $A$ and $B$ nonzero, while $a=2\pi/N$ and
$b=2\pi /M$ \cite{SUSYProductGaugeGroup}. Minima at large $X$ are
then generic for large values of $N$ and $M$, with $M$ close to
$N$.

In terms of the real component fields the supersymmetric part of
the potential obtained using the superpotential (\ref{RacetrackW})
takes the following form (for real $W_0$):
\beqa\label{potencial1}
    V_F & = & \frac{e^{-(a+b)X}}{6X^2} \Bigl\{ aA^2
    \left(aX+3\right)~ e^{(b-a)X} +  bB^2\left(bX+3\right)
    \, e^{-(b-a)X} \nonumber\\
    && \qquad\qquad + AB\left(2abX + 3a +3b\right) \cos
    [\left(a-b\right)Y]  \\
    &&\qquad\qquad\qquad + 3 W_0 \left(aA e^{bX}  \cos [aY]
    + bBe^{aX}  \cos [bY ]\right) \Bigr\} \nonumber
\eeqa
The scalar potential obtained by summing this with $\delta V =
E/X^\alpha$ has several de Sitter minima, depending on the values
of the parameters $A,a,B,b,W_0$ and $E$. In general the different
periodicities of the $Y$-dependent terms lead to a very rich
landscape of vacua \cite{egq, bkl}. This is particularly so if
$a-b$ is taken to be very small (as is done for the standard
racetrack models), such as if we take $a = 2\pi /M$ and $b =
2\pi/N$ with integers $N\sim M$ and both large.

The pattern of vacua obtained using (\ref{RacetrackW}) differs
from that found with the original KKLT scenario in that it allows
nontrivial minima even when $W_0=0$. For $W_0\neq 0$, many new
local minima appear due to the small periodicity of the terms
proportional to $W_0$ in the scalar potential. By choosing the
background fluxes appropriately we have the freedom to adjust the
values of $W_0$ and $E$. In particular $E$ can be tuned, as in
KKLT, to adjust the value of the potential at its minimum. Unlike
for the KKLT case, these parameters can also be adjusted to
arrange the existence of local minima which are both
supersymmetric and flat in four dimensions \cite{bkl}.

\subsection{Inflationary Slow Roll}

Another important difference between the superpotentials
(\ref{KKLTW}) and (\ref{RacetrackW}) is seen once one examines the
dynamics of the field $T$ as it rolls towards these vacua. In
particular, for inflationary purposes our interest is in slow
rolls, for which the scalar-field energy is dominated by its
potential rather than its kinetic energy. The main point of
ref.~\cite{racetrack} was that such slow rolls can exist for the
superpotential (\ref{RacetrackW}), even though they do not for
(\ref{KKLTW}).

A sufficient condition for the occurrence of an inflationary slow
roll is $\epsilon, |\eta| \ll 1$, where $\epsilon$ and $\eta$ are
the slow-roll parameters \cite{Revs}, suitably generalized
\cite{bcsq,SRnonmin} for scalars having nonminimal kinetic terms,
${\cal L}_{\rm kin} = - \frac12 \, g_{ab}(\phi) \partial_\mu
\phi^a
\partial^\mu \phi^b$. Explicitly, these are
\beq
    \epsilon = \frac12 \left( \frac{g^{ab} \, \partial_a V
    \partial_{b} V}{V^2} \right)
\eeq
while $\eta$ is defined as the most-negative eigenvalue of the
matrix
\beq
    {N^a}_b = \left[ \frac{g^{ac}(
    \partial_c \partial_b V - \Gamma^d_{cb} \partial_d V)}{V} \right]
     \,,
\eeq
These definitions assume Planck units, and in them the indices
`$a,b,c,d$' run over a complete basis of fields, $\phi^a$. The
connection, $\Gamma^d_{cb}(\phi)$, appearing in $\eta$ is
constructed in the usual way from the target-space metric,
$g_{ab}(\phi)$, which is in turn defined by the scalar-field
kinetic terms. The point behind these definitions is their
invariance under redefinitions of the scalar fields, and their
reduction to the standard ones when evaluated for fields with
minimal kinetic terms.

Notice also that the complications in $\eta$ having to do with the connection
$\Gamma^a_{bc}$ are irrelevant when specialized to points for
which $\epsilon$ vanishes.

The above expressions simplify when expressed in a complex field
basis $\{ \phi^a \} = \{ \varphi^i, \overline\varphi^{\bar\jmath}
\}$, for which the target-space metric has components $g_{ij} =
g_{\bar\imath\bar\jmath} = 0$ and $g_{i\bar\jmath} = g_{\bar\imath
j} = \partial_i \partial_{\bar\jmath} K$. In this case the only
nonzero connection components are $\Gamma^i_{jk} = K^{i\bar{n}}
\partial_j \partial_k \partial_{\bar{n}} K$ and their complex
conjugates. The above definitions then reduce to
\beq
    \epsilon = \left( \frac{K^{i\bar\jmath} \, \partial_i V
    \partial_{\bar\jmath} V}{V^2} \right)
\eeq
and
\beq
    {N^a}_b
    = \left( {\begin{array}{cc} {N^i}_j & {N^i}_{\bar\jmath} \\
    {N^{\bar\imath}}_j & {N^{\bar\imath}}_{\bar\jmath}
    \end{array}}
    \right) \,,
\eeq
where
\beq
    {N^i}_k = \frac{K^{i\bar\jmath} \partial_{\bar\jmath} \partial_k
    V}{V} \,, \qquad
    {N^k}_{\bar\imath} = \frac{K^{k\bar\jmath} (
    \partial_{\bar\imath} \partial_{\bar\jmath} V - K^{l\bar{n}}
    \partial_{\bar\imath} \partial_{\bar\jmath} \partial_l K \,
    \partial_{\bar{n}} V)}{V} \,,
\label{srgen}
\eeq
and ${N^{\bar\imath}}_{\bar\jmath}$ and ${N^{\bar{k}}}_i$ are
obtained from these by complex conjugation.

Usually a fine tuning of some of the parameters of the model are
required in order to ensure that both $\epsilon$ and $|\eta|$ are
sufficiently small. For instance for the superpotential
(\ref{RacetrackW}) it was found in \cite{racetrack} that the
potential has a saddle point (for which $\epsilon = 0$) near a
flat local minimum, but a tuning of the order of 1 part in 1000
was needed, in some of the parameters
of the model, in order for $|\eta|$ to be small enough to obtain the
minimal $60$ $e$-foldings of inflation ( See \cite{racetrack} for details.).

Once an inflationary region is obtained, the observed size of the
CMB temperature fluctuations may be obtained by adjusting the
string scale, typically leading to a value close to the GUT scale.
The consistency of this with the observed value of Newton's
constant then implies a condition on the VEV of the volume modulus
that determines the string scale from the Planck scale. Even
though this procedure looks (and is) quite restrictive, solutions
nonetheless exist having acceptable values for all of these
experimentally measurable quantities. In
refs.~\cite{bcsq,racetrack} the scaling properties of the
low-energy action were exploited in finding values of the
parameters which satisfied all of these criteria.

\section{The orientifold of $\IP^4_{[1,1,1,6,9]}$ }\label{example}

We now return to the main line of development, and repeat the above
steps for the orientifold of 
degree 18 hypersurface 
$\IP^4_{[1,1,1,6,9]}$, an elliptically fibered Calabi-Yau over
$\mathbb{P}^2$. The stabilization of moduli in this model was performed 
in \cite{ddf} where it was also shown how D3 instantons
generate a  nonperturbative superpotential, thus providing an explicit
realization of the KKLT scenario.\footnote{Furthermore, this model 
has been used as the prototype  for a general class of models in 
which the $\alpha'$ corrections to the K\"ahler potential give rise 
to exponentially large volume compactifications \cite{bbcq}. For 
simplicity we restrict ourselves here to the leading-order potential, 
and argue that our analysis can easily be extended to the exponentially 
large volume case since in the end it is the axionic field that plays 
the role of the inflaton and not the volume. We nevertheless verified that, 
in the range of parameters that we are working on, the $\alpha'$ corrections 
do not affect the results substantially and can be safely ignored.}

The model is a Calabi-Yau threefold with the number of K\"ahler
moduli $h^{1,1}=2$ and the number of 
complex structure moduli $h^{2,1}=272$. The 272 parameter prepotential
for this model is not known. 
We will restrict ourselves to the slice of the complex structure
moduli space which is fixed under the
 action of the  discrete symmetry $\Gamma\equiv \mathbb{Z}_6 \times
 \mathbb{Z}_{18}$. This allows to 
reduce the  the moduli  space of the complex Calabi-Yau structures
to just 2 parameters, since the 
slice is two-dimensional. This restricted model has a long string
pedigree, starting with
 \cite{Candelas:1994hw};  it is a hypersurface in the weighted projective space
 $\IP^4_{[1,1,1,6,9]}$. The remaining 
270 moduli are required to vanish to support this symmetry. The
defining equation for the 
Calabi-Yau 2-parameter subspace of the total moduli space is \beq
 f=   x_1^{18} + x_2^{18}+ x_3^{18}+ x_4^3 + x_5^2 - 18 \psi x_1 x_2
x_3 x_4 x_5 - 3 \phi x_1^6 x_2^6 x_3^6 \,.
\label{CY}\eeq

The first stage of the GKP-KKLT scenario \cite{gkp,kklt}, stabilization of 
the type IIB axion-dilaton $\tau$ and the two complex structure 
moduli $\psi$ and $\phi$  in eq. (\ref{CY}) was performed in \cite{ddf} 
explicitly. It was important at this stage to turn on only the fluxes 
on $\Gamma$-invariant cycles, the same tool has been used in other 
models of flux vacua stabilization in \cite{Giryavets:2003vd}.

The K\"ahler geometry of the remaining two K\"ahler  moduli $h^{1,1}=2$  
was specified in  \cite{Candelas:1994hw,ddf}.  We denote them 
by $\tau_{1,2}=X_{1,2} +i Y_{1,2}$. These moduli correspond 
geometrically to the complexified volumes of the divisors 
(or four-cycles) $D_4$ and $D_5$, and give rise to the gauge 
couplings for the field theories on the D7 branes which wrap 
these cycles.  For this manifold the K\"ahler potential  is
given by
\beq\label{Kahler}
   K = - 2 \ln R \,,
\eeq
where $R$ denotes the volume of the underlying Calabi-Yau space,
given in terms of the two K\"ahler moduli by
\beq\label{volume}
   R = \frac{\sqrt{2}}{18}\left({X_2}^{3/2} - {X_1}^{3/2}\right)
   \,.
\eeq
As is easily verified, this K\"ahler potential satisfies the
identity $K^{i\bar\jmath} K_i K_{\bar\jmath} = 3$, and so is of
the no-scale type, showing that both $\tau_1$ and $\tau_2$
represent flat directions so long as the superpotential does not
depend on them (as is true in particular for the Gukov-Vafa-Witten
superpotential \cite{gvw} for these fields). These flat directions
are lifted by D3 instantons, which have been computed for this
manifold \cite{ddf} to generate the following nonperturbative
superpotential:
\beq\label{dosexp}
   W=W_0+A\,e^{-a\tau_1}+B\,e^{-b\tau_2}\,.
\eeq
This form is similar to the racetrack models inasmuch as it
involves two exponential terms, but differs in that each
exponential depends only on one of the two complex K\"ahler moduli.

Given these expressions for $K$ and $W$, the supersymmetric part
of the scalar potential takes the following form:
\beqa \label{scalarpot} V_F &= \frac{108}{
({X_2}^{3/2}-{X_1}^{3/2})^2} \big\{4 (X_1X_2)^{1/2} |X_2^{1/2}
  W_{\tau_1} + X_1^{1/2} W_{\tau_2}|^2\cr
&  -3 X_1({W_{\tau_1}}^*W+W_{\tau_1} W^*)-3
X_2({W_{\tau_2}}^*W+W^*W_{\tau_2})\cr & +2|X_1 W_{\tau_1}+ X_2
W_{\tau_2}|^2\big\}\,, \eeqa
which gives the following function of $X_i$ and $Y_i$:
\newpage
\beqa \label{scalarpot2}
    V_F &={216 \over
    ({X_2}^{3/2}-{X_1}^{3/2})^2}\big\{ B^2 b(b {X_2}^2 + 2
    b {X_1}^{3/2} {X_2}^{1/2} + 3 X_2) e^{-2 b X_2}\cr &
    + A^2 a(3 X_1 + 2 a {X_2}^{3/2} {X_1}^{1/2} + a X_1^2)
    e^{-2 a X_1}\cr & +3 B b \, W_0 X_2 e^{-b X_2} \cos(b Y_2) + 3 A a\, W_0  X_1
    e^{-a X_1} \cos(a Y_1)\cr & +3 AB e^{-a X_1 -b X_2} (a X_1
    +b X_2 + 2 a b X_1 X_2) \cos(-a Y_1 + b Y_2))\big\}\,
\eeqa
Notice that this potential is parity invariant, $(X_i,Y_i) \to
(X_i,-Y_i)$, with $Y_i$ being pseudoscalars. It is also invariant
under the two discrete shifts, $Y_1 \to Y_1 + 2\pi m_1/a$ and $Y_2
\to Y_2 + 2\pi m_2/b$, where the $m_i$ are arbitrary integers.
Notice also the approximate $U_R(1)$ $R$-symmetry, $a\delta Y_1 =
b\delta Y_2 = \epsilon$, which becomes exact in the limit $W_0 \to
0$. Since the $Y_i$ only enter the full potential through $V_F$, these
fields may be fixed without reference to the nonsupersymmetric
part of the potential, $\delta V$. If $a,b,A,B$ and $W_0$ are all
positive (as we assume for simplicity) then inspection of
eq.~(\ref{scalarpot2}) shows that for fixed $X_i > 0$, the
potential $V_F$ would be smallest if we could choose $\cos(aY_1) =
\cos(bY_2) = \cos(-aY_1 + bY_2) = -1$ (or maximized by choosing
them equal to $+1$). But this cannot be done since these three
conditions are mutually incompatible. For small $W_0$, the case of
interest in what follows, it is energetically preferable to have
$\cos(-aY_1 + bY_2) = -1$, leading to
\beqa
    a Y_1 - b Y_2 = \pi \qquad \hbox{(mod $2\pi$)} \,,
\eeqa
and to allow $\cos(aY_1)$ and/or $\cos(bY_2)$ to be larger than
$-1$. For instance, if we follow Douglas {\it et.~al.}~\cite{ddf}
by making the choices $A=B=1$, $a=2 \pi /4$ and $b= 2 \pi /30$
then because $a > b$ the minima prefer $\cos(bY_2) = \cos(-aY_1 +
bY_2) = -1$, corresponding to the following lattice of degenerate
minima: $(Y_1, Y_2) = (4m_1,30 m_2 - 15)$, with $m_1$ and $m_2$
integers.

We find numerically that using these values for $Y_i$ in $V_F$
leads to a unique minimum for $(X_1,X_2)$. The precise position of
this minimum varies with the parameters $A,B,a,b$ and $W_0$, and
in particular the minimum is shifted to arbitrarily large values
of $X_i$ as $W_0 \rightarrow 0$. The potential $V_F$ is negative
when evaluated at this minimum, but it can be made positive once
we add the anti-D3 branes \`a la KKLT. For this purpose we take
$\delta V = {E / R^2}$ if warping is not important at the anti-D3
position, or $\delta V = {E / R^{4/3}}$ if warping is important.
In both cases $E$ is a positive constant, as in previous sections.
For our subsequent numerical purposes we use the unwarped
case in what follows, and write the lifting term as
\beq
    \delta V = {D\over \left(X_2^{3/2} - X_1^{3/2}\right)^2}
\eeq

\section{Inflationary parameters and slow roll}\label{roll}

We next ask whether slow-roll evolution is possible with this
superpotential and K\"ahler potential. We are guaranteed the
existence of saddle points for which $\epsilon = 0$ because of the
existence of periodic minima in the $(Y_1,Y_2)$ plane. For
instance, when $a \gg b$ --- like for the parameters chosen in
ref.~\cite{ddf} --- we found above that the minima correspond to
the choices $aY_1 = 0$ (mod $2\pi$) and $bY_2 = \pi$ (mod $2\pi$).
Saddle points are then found midway in between, such as for $aY_1 =
bY_2 = \pi$ (mod $2\pi$). The question is whether the parameters
$a,b,A,B$ and $W_0$ can be chosen to ensure $|\eta|$ is small
enough at these points.

Numerically, we find the following results. For small $W_0$ the
minimum discussed above exists for large $X_{1,2}$, corresponding
to large volume $R$. 
 Fixing the $X_i$ fields at their particular values
of the global minimum of the potential, we can find
an infinite number of maxima and minima obtained from
one another by shifts in the two $Y_i$ directions.
 There are saddle points in between these
minima, at $(aY_1,bY_2) = (\pi+2\pi n,\pi+2\pi n)$,
whose unstable directions lie purely within the
$(Y_1,Y_2)$ subspace of the four possible field directions. 
Since these are purely axionic directions, the inflation we obtain
resembles in some ways the natural inflation mechanism
\cite{nat}. However, the scalar potential and the field 
evolution during inflation in our model differ significantly 
from their counterparts in the natural inflation scenario.

Because of the approximate $U_R(1)$ symmetry which appears in the
limit $W_0 \to 0$, the flatness of the potential at these saddle
points turns out to depend sensitively on the value of $W_0$. For
$W_0=0$ one of the $Y_i$ directions would be perfectly flat,
corresponding to the direction $aY_1 \propto bY_2$ which is the
Goldstone boson for this symmetry.  This can be seen by expanding
the potential near a saddle point,
$aY_1 =\pi + y_1,\ bY_2 = \pi+y_2$:
\beqa
V &\sim& f(X_1,X_2) +
\,{ A}\,a{ W_0}\,{ X_1}\,{e^{-a{ X_1}}}{{ y_1}}^{2}
+\,{ B}\,b{ W_0}\,{ X_2}\,{e^{-b{ X_2}}}{{ y_2}}^{2}\nonumber\\
&-&\,{
 A}\,{ B}\,{e^{-a{ X_1}-b{ X_2}}}\left( a{ X_1}
+b{
X_2}+2\,ab{ X_1}\,{ X_2} \right)  \left( { y_1}-{ y_2} \right)
^{2}
\eeqa
However, for $A,B>0$, the direction orthogonal to $y_1=y_2$ is
unstable, so $W_0\to 0$ is not the tuning needed to get a nearly
flat saddle point.  Rather we need to tune the $W_0$-dependent
terms against those which are independent of $W_0$.
Moreover, unlike the single-modulus case of the previous
section (but similar to KKLT), in the limit $W_0 = 0$ there
is no minimum in the $X_{1,2}$ directions. Both of these
considerations show that the optimal value of $W_0$ is nonzero.
For fixed values of $a,b,A,B$ (and adjusting $D$ so as to keep
the final vacuum energy zero), the saddle point at
$(aY_1,bY_2) = (\pi,\pi)$ becomes increasingly
flat as $W_0$ is increased, up to some critical value
$W_c$ beyond which
it is no longer a saddle point, but becomes a shallow local minimum.

Searching the parameter space,  we are able to find choices
for which the scalar potential behaves similarly to
the original racetrack inflation potential.  Starting at the
saddle point,
since only one of the four real directions is unstable, we
have sufficient freedom to make this direction
flat enough to give rise to successful inflation. The technical
details are more complicated in this case than for the
original racetrack model, as we now show.

Our goal was to find a set of parameters which would lead to 
inflation satisfying the COBE normalization of power spectrum,
\beq
    P(k_0) = 4\times 10^{-10}
\eeq at the scale $k_0/(a_0 H_0) = 7.5$.  If there are $N_e$
$e$-foldings of inflation after horizon crossing, this corresponds
to $\Delta N = N_e - \ln(7.5) = N_e-2$ $e$-foldings before the end
of inflation. We also need to have the spectrum to be sufficiently
flat, $n_{s} = 0.95 \pm 0.02$ \cite{MacTavish:2005yk}.

After some searching of parameter space, we found a few examples
which satisfy these criteria. These examples are not particularly 
easy to find.  The example with $P(k_0) = 4\times 10^{-10}$ and 
$n_{s} = 0.95$, on which we focus for the rest of the paper, 
has the following parameters:
\beqa\label{values}
    W_0&=&  5.22666 \times 10^{{-6}},\quad
    A=0.56, \quad B=7.46666 \times 10^{{-5}},\quad
    a=2\pi/40, \quad b= 2\pi/258,\nonumber\\
 D &=& 6.21019\times 10^{-9}
\eeqa
where $W_0$ is chosen to be close to the critical value
mentioned above; hence these are optimal values for getting
a long period of inflation and a flat spectrum of density
perturbations.

The choice of parameters $W_0$, $A$,  $a$  and $D$ in
eq. (\ref{values}) is quite reasonable from the point of view of
already available  stringy construction in \cite{ddf}. The situation
is  more tense for our choice of $B$ and $b$. It may be difficult to
get such a small value of $B$ but taking into account of the fact that
the value of $B$ depends on the stabilization point for complex 
structure moduli, it does not seem to be impossible.
On the other hand, to find explicit stringy constructions with a large
value of the inverse
 of $b$  may require special
effort.
 Since we take care that the volume
 of stabilization
 with such parameters is still large in stringy units, we conclude
 that the full 
set of parameters in eq. (\ref{scalarpot2}) is possible in principle,
however, 
the extreme values of $B$ and $b$ may need a better justification in
more general
 explicit constructions.

\DOUBLEFIGURE[ht]{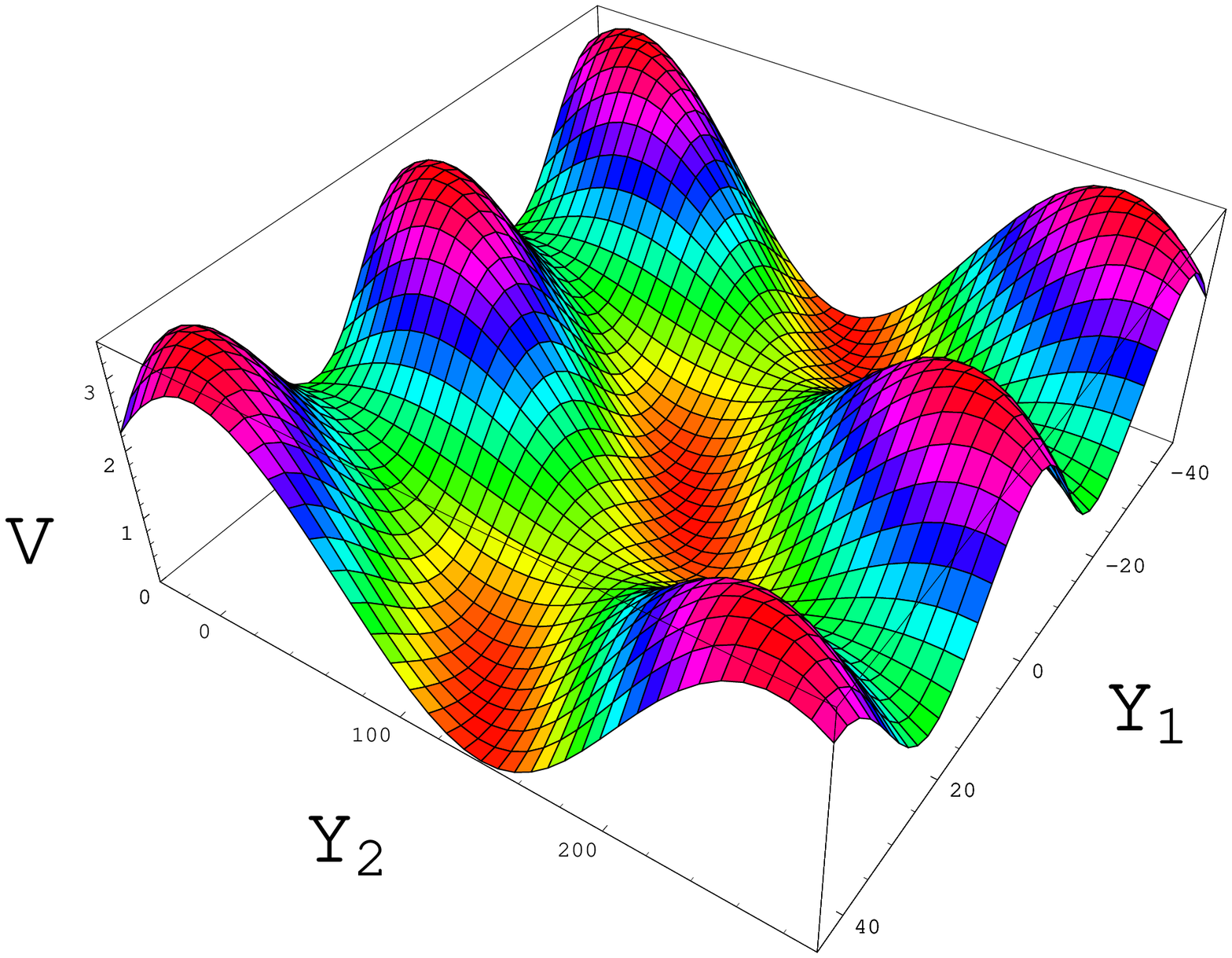, width=1\hsize}{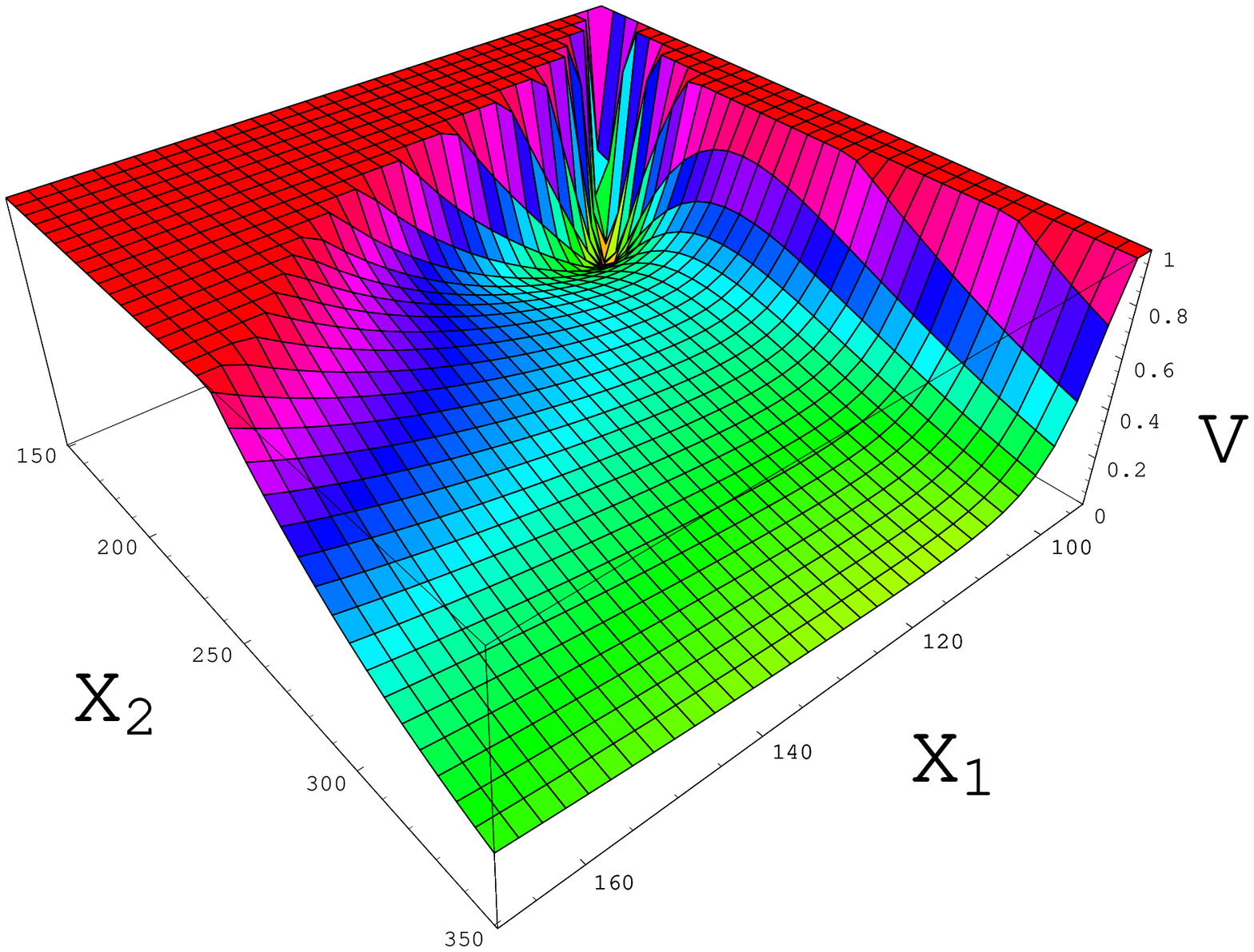,
  width=1\hsize}{The potential as a function of the axion variables $Y_1$, $Y_2$
  at the minimum of the radial variables $X_1, X_2$, , in units $10^{{-15}}$ 
of the Planck density.}{The potential as a function of the radial variables $X_1$, $X_2$
  at the minimum of the angular variables $Y_1, Y_2$, in units $10^{{-14}}$ of the Planck density.}

With these choices of the parameters the minimum described above is located at
\beq\label{min}
    X_1=98.75839 \qquad X_2=171.06117\qquad Y_1=0\qquad
    Y_2= 129 \,,
\eeq
corresponding to a volume $R=99$ in string units, which is
large enough to trust the effective field theory treatment we use.
The parameter $D$ is tuned so that the
potential vanishes at this minimum.

It is very difficult to plot the  potential since it is a function 
of 4 variables. Here we will only show the behavior of this potential 
as a function of the axion variables $Y_1$, $Y_2$  at the minimum of 
the radial variables $X_1, X_2$, and the potential as a function of 
the radial variables $X_1$, $X_2$ at the minimum of the angular 
variables $Y_1, Y_2$. Figures 1 and 2 illustrate the behavior of 
the potential near the minimum (\ref{min}).

We have
checked that the eigenvalues of the Hessian (mass$^2$) matrix are
all positive, verifying that it is indeed a local minimum. The
value of the masses for the moduli at this minimum turn out
to be of order $10^{-6} - 10^{-7}$ in Planck units.

Inflation occurs near the saddle point  located at
\beq
    X_1=108.96194\qquad X_2=217.68875\qquad
    Y_1=20\qquad Y_2=129 \,.
\eeq

At this point the mass matrix has three positive eigenvalues and
one negative one in the direction of $(\delta X_1, \delta X_2,
\delta Y_1, \delta Y_2) = (0, 0, -0.6546, 0.7560)$, corresponding
to a purely axion direction. This is the initial direction of the
slow roll away from the saddle point towards the nontrivial
minimum described above.

\EPSFIGURE[r]{hump.eps, width=8cm}{A cross section of the
change in the potential near the saddle point, along the tangent
to the initial unstable direction, in units of the Planck density.
\label{hump} }

The value of the effective potential at the saddle point is $V \sim
3.35 \times 10^{{-16}} = M^{4}$ in Planck
units, so that the scale of inflation is $M =
3.25\times 10^{14}$ GeV. This is a rather small scale. The ratio of 
tensor to scalar perturbations in this scenario is very small, 
$r\ll 1$, so the gravitational waves produced in this scenario will be very
hard to observe  \cite{Efstathiou:2006ak}.  We plot the shape of
the potential along the initial unstable direction, $(Y_1,Y_2) =
(-0.6546, 0.7560)\phi$ in figure \ref{hump}.  (The zero of the
$y$-axis is not the true minimum of the potential since the real
inflaton trajectory curves away from the initial tangent to the
trajectory.)

To find the slow roll parameter $\eta$ at the saddle point
(recall that $\epsilon = 0$ automatically at a saddle point), as well as to compute
the inflationary trajectories, we must also specify the kinetic
terms for the fields.  The general supergravity expression is
given in terms of derivatives with respect to the K\"ahler potential,
\beq
    {\cal L}_{\rm kin} = {\partial^2 K\over
    \partial \phi_i^{*}\partial \phi_j} \,
\partial_\mu\phi_i^{*}\partial^\mu \phi_j
\eeq
Explicitly, we find
\beqa
    {\cal L}_{\rm kin}\ &=&\
    \frac{3}{8\left(X_1^{3/2}-X_2^{3/2}\right)^2}
    \left(\frac{2X_1^{3/2}+X_2^{3/2}}{\sqrt{X_1}}\
    \left(\partial X_1^2+\partial Y_1^2\right)\ \right.
 \nonumber \\ \ &-& 6\sqrt{X_1X_2}\,
    \left(\partial X_1\partial X_2 + \partial Y_1 \partial
Y_2\right)
    +\left.
    \frac{X_1^{3/2}+2X_2^{3/2}}{\sqrt{X_2}}\,
    (\partial X_2^2 + \partial Y_2^2)
    \right)
\eeqa

The
noncanonical kinetic terms require the use of the generalized
definitions of the slow-roll parameters defined earlier.
Alternatively, because the initial roll is purely in the $(Y_1,
Y_2)$ plane we can instead choose to diagonalize the kinetic terms
for the $Y_i$ fields in the approximation that the $X_i$'s are
constants.   This diagonalization
is straightforward but cumbersome, and
leads to the value
\beq
    \eta = -0.01
\eeq at the saddle point.  As we will discuss below, this leads to
$n_{s} \approx 0.95$ and a long period of inflation, 980
$e$-foldings after the end of eternal inflation.

\section{Normalization of spectrum and scale of inflation}\label{spectrum}

We have computed the power spectrum for the model under consideration
by first
numerically evolving the full set of field equations, which can
be efficiently written in the form
\beqa
{d\phi_i\over dN} &=& {1\over H} {\dot\phi_i}(\pi_i) \nonumber\\
{d\pi_i\over dN} &=& -3\pi_i -{1\over H} {\partial\over
\partial\phi_i}\left(V(\phi_i) - {\cal L}_{\rm kin}\right)
\eeqa where $N$ is the number of $e$-foldings starting from the
beginning of inflation, $\pi_i =
\partial{\cal L}_{\rm kin}/\partial\dot\phi_i$ are the canonical
momenta, and the time derivatives ${\dot\phi_i}$ are regarded as
functions of $\pi_i$, which can conveniently be solved for using
symbolic manipulation.  (This procedure alleviates the need for
computing Christoffel symbols in field space, or explicitly
diagonalizing the kinetic term.  However we have also checked our
results by directly integrating the second order equations for
$X_i, Y_i$ in Mathematica.)   We use initial conditions where the field
starts from rest along the unstable direction, close
enough to the saddle point to give more than 60 e-foldings
of inflation. In fact our starting point corresponds to the
boundary of the eternally inflating region around the
saddle point where the field is dominated by quantum
fluctuations \cite{topinf} \footnote{This realization of eternal
  inflation was also 
used in \cite{racetrack, graham} to argue that the standard overshoot problem
of string cosmology \cite{overshoot} is ameliorated. In \cite{graham}
it was further argued that this set-up relaxes the late-thermal roll
problem in which temperature corrections to the effective potential
may destabilize the vacuum \cite{thermal}.}.
  An example of the inflationary trajectories for all the 
fields is shown in figures \ref{y-evolution}-\ref{x-evolution}.

\DOUBLEFIGURE[h]{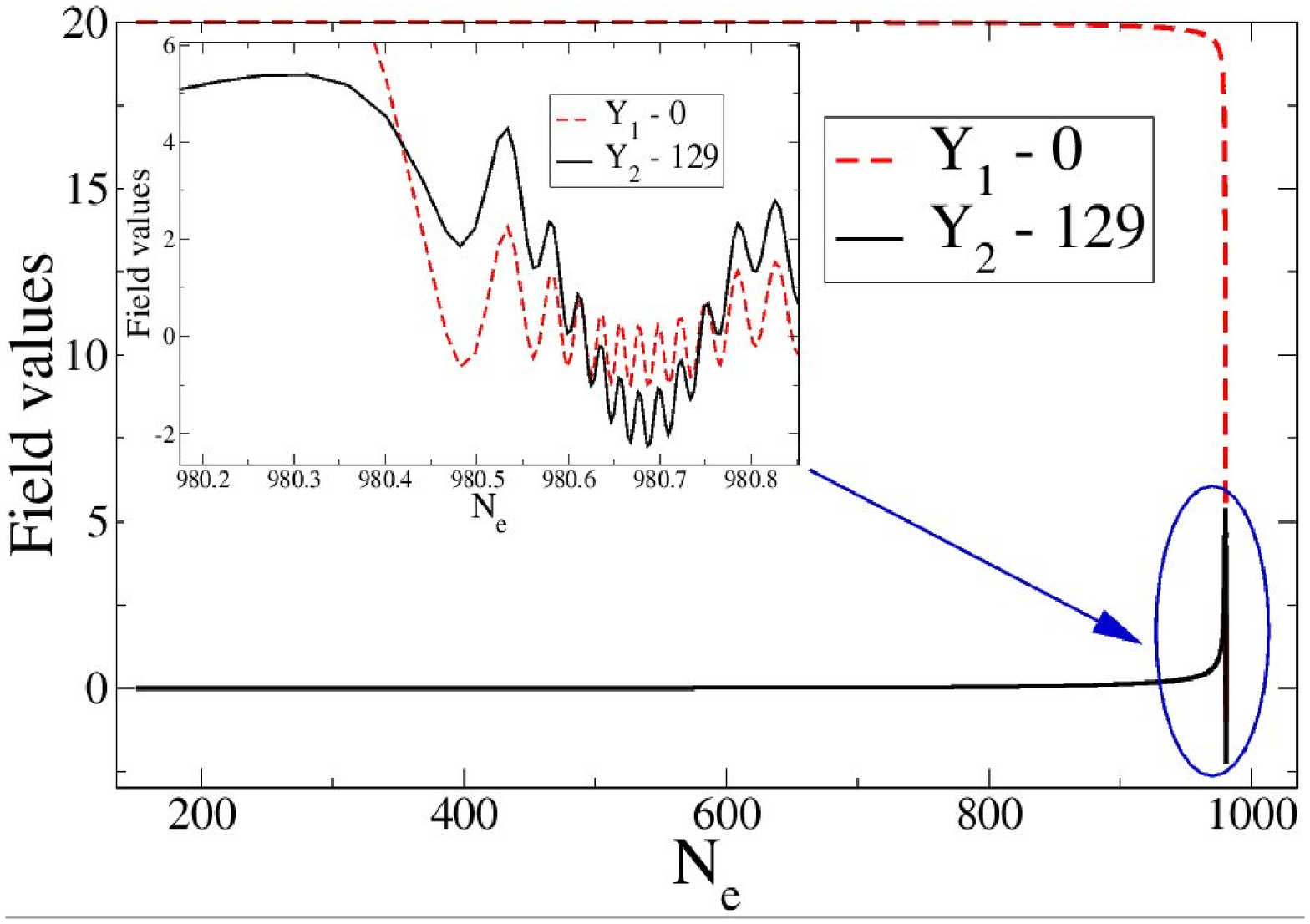, width=8.cm}{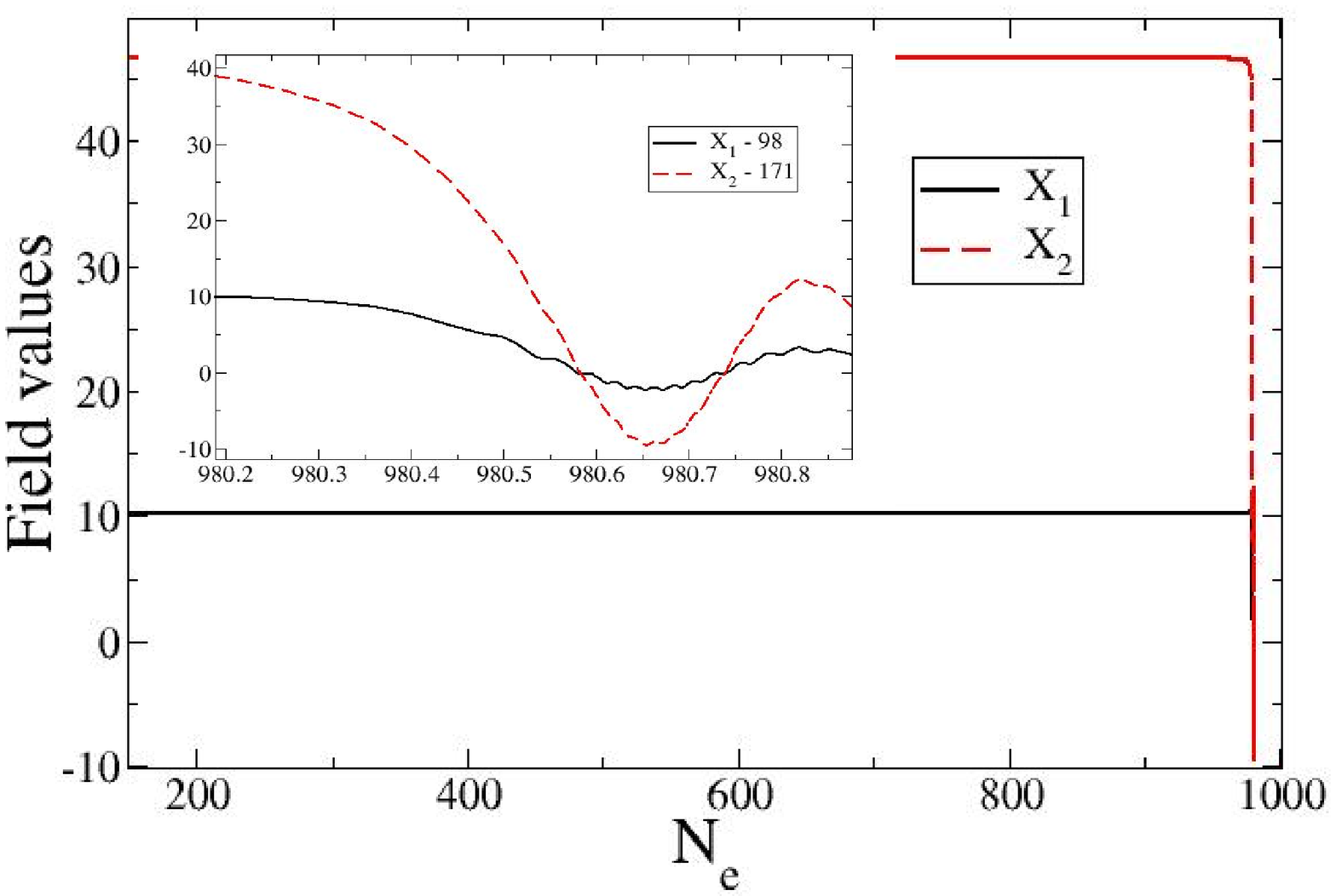, width=8.7cm} 
{Evolution of the axionic fields $Y_1,Y_2$ during inflation. 
Inset shows oscillations around the minimum at the end of 
inflation.\label{y-evolution}}{Evolution of the $X_1,X_2$ directions
during the inflationary period. Note that even though
the potential at the minimum is only protected by a small barrier, 
our numerical evolutions shows that the fields do relax to this 
minimum evading a potential overshooting problem.\label{x-evolution}}

To compute the spectrum of adiabatic scalar density perturbations,
we  use the effective
one-field approximation,
namely that the power is given by
\beq
\label{power}
    P(k) = {1\over 50\pi^2} {H^4\over {\cal L}_{\rm kin}}
\eeq evaluated at horizon crossing, $k/a = H$.  Equivalently, we
can parametrize $P$ as a function of the number of $e$-foldings of
inflation.

 If we assume that the
reheat temperature is of the same order as the scale of
inflation,  $T_{r }\sim M = 3.25\times 10^{14}$ GeV, the number of
$e$-foldings of inflation since horizon crossing is \beq
    N_e = 53 + \ln(M/10^{13}{\rm \ GeV}) \approx 56.5
\eeq and the COBE normalization scale is at $\Delta N = N_{e}- \ln
(k_{0} /(a H_{0} )) = N_e - \ln(7.5) = N_e-2 \sim 55$ $e$-foldings
before the end of inflation.   If $T_r < M$, then $N_e$ is reduced
by the amount $\frac13\ln(M/T_r)$. For example, one might like to
satisfy the gravitino bound $T_r < 10^{10}$ GeV, which decreases
$N_e$ to 53.  The running of the spectral index (which will be
studied in the next section) is sufficiently small that this
changes the value of $n_s$ only by $-0.003$, thus our results do
not depend sensitively on assumptions about the reheating
temperature.

Our numerical analysis shows that the amplitude of density 
perturbations for the parameters given above does satisfy 
the COBE normalization of the power spectrum,
\beq
    P(k_0) = 4\times 10^{-10}
\eeq
at the COBE normalization scale.

\section{Power spectrum, dependence on $W_0$, and anthropic 
considerations}\label{ns}

The choice of the parameters leading to $P(k_0) = 4\times 10^{-10}$ 
is not unique. First of all, just as in the Racetrack scenario
\cite{racetrack},  there is a rescaling of parameters
\beq
\label{scaling}
(a,b)\to \lambda^{-1}\, (a,b), \quad 
\left(A,B, W_0,D^{1/2}\right)\to \lambda^{3/2}\left(A,B, W_0,D^{{1/2}}\right), 
\quad \left(X_i, Y_i\right)\to \lambda\ \left(X_i, Y_i\right)
\eeq

This transformation does not alter inflationary dynamics or the
height of the potential; it rescales the fields, but leaves the
slow-roll parameters and the amplitude of density perturbations
invariant. If a set of parameters gives rise to slow-roll inflation
in a region of field space, the transformed parameters will also yield
inflation with the same spectrum of density fluctuations, in
the transformed region of field space.

A less trivial change occurs if we keep $a,b,X_i, Y_i$ invariant and
rescale $W_0, A, B$ by $\mu^{-3/2}$ each and $D$ by $\mu^{-3}$. This
transformation rescales the potential $V$ and the amplitude of
density perturbations $P$ as $\mu^{3}$  without altering the values
of the fields, the slow roll parameters, or the total duration of
inflation.

It would be interesting to compare different sets of parameters
related to each other by the $\mu$-transformation and  find  which of
these parameters are more probable in the stringy landscape, which 
ones lead to the greater volume for the inflationary universe and more
efficient reheating, {\it etc}. This would allow us to determine the most
probable amplitude of density perturbations in this class of
theories, see {\it e.g.} \cite{Garcia-Bellido:1993wn} for a discussion of
closely related issues. This is a complicated problem which goes
beyond the scope of the present investigation. We still do not know what is the
proper choice for the probability measure in eternal
inflation. In addition, it is not obvious whether  a simultaneous
scaling of several different parameters which have different origin
can be easily achieved within the full string theory landscape.

Therefore we may instead pursue a more modest goal and study what happens
when we change just one of the parameters, {\it e.g.,} $W_{0}$. Let
us recall the arguments concerning the distribution of flux vacua in
string theory with a given value of $W_0$, \cite{Shamit},
\cite{Douglas:2004qg}. This distribution is believed to be uniform
near zero. This means that  if one wants $|W_0| \leq \epsilon$, the
fraction of flux vacua that may provide this value is of the order
$\epsilon^2$. Keeping in mind an enormously  large number of flux
vacua we deduce that any value of $W_0$ which we may need for
cosmology is available, at least in principle, in some explicit
constructions, in particular in the ``better racetrack'' model. To
obtain the particular value which we need would require an
intensive numerical search, by
varying sets of stabilizing fluxes, and this is not
guaranteed to be a computable problem in a reasonable amount of time
\cite{Denef:2006ad}. It is therefore satisfying to know that at least
in principle, the value of $W_0$ which we need in our model for
cosmology is possible. Particularly, we may  relax the restriction to
the 2-parameter model defined in eq. (\ref{CY}) and engage all 272
complex structure moduli of the full model, which admits a huge
number of possible fluxes. In this way we expect that
eventually any required value of $W_0$
can be achieved constructively.

Since inflation in our scenario requires fine tuning, it is not
surprising that a change of $W_{0}$ by several percent can spoil
inflation. We have found, for example, that when one decreases
$W_{0}$ from $5.227\times 10^{-6}$ to $5.147 \times 10^{-6}$, i.e. by
1.5\%, the height of the saddle point and the amplitude of
perturbations change insignificantly, while $n_s$ decreases
to $0.92$, which is at the verge of being ruled out by
observations.

We have evaluated the scalar spectral index, $n_s = d\ln P/d\ln k$
at two different points along the inflationary trajectory: at the
beginning, when the fields are  near the saddle point, and at 50
$e$-foldings before the end of inflation, near the COBE
normalization point.  We carried this out for a range of $W_0$
values around our fiducial value $W_0 = 5.227\times 10^{-6}$,
going up to the critical value $W_c = 5.267\times 10^{-6}$, beyond
which the saddle point becomes a local minimum. The results are
shown in figure \ref{index}.

\DOUBLEFIGURE[h]{index5.eps, width=8.5cm}{pert-seq2.eps,
width=7.9cm} {Spectral index at the saddle point and at  the COBE
scale, as a function of $W_0$ divided by its optimal value, with
other parameters (except $D$) held fixed. \label{index}}{Spectral
index as a function of the number of $e$-foldings of inflation
(minus the total number of $e$-foldings), for the original
racetrack model, and the two-K\"ahler modulus model of this paper.
\label{comp}}

Evaluating the spectral index at $55$ $e$-foldings before the end
of inflation gives the spectral properties relevant for the CMB.
Figure \ref{index} shows that for $W_0 = 5.227\times 10^{-6}$ the
spectral index reaches its largest value \beq
    n_s  \approx  0.95
\eeq This is the same value which we found in the original
racetrack model. The figure shows that $W_0$ has to be tuned at
the level of a percent to keep the spectral index from decreasing
into a range of phenomenologically disfavored values. The
comparison between the  model of this paper and the original
single-K\"ahler modulus racetrack model is shown in figure
\ref{comp}, where we chose the endpoint of inflation to coincide
for the two models, for ease of comparison. The figure plots the
spectral index as a function the number of $e$-foldings, minus the
total number of $e$-foldings, showing that the spectral properties
of the two models are remarkably similar.

During most of inflation, the parameter $n_s$ remains nearly
constant, at its value near the saddle point, shown in figure
(\ref{index}). Only during the latter part of its evolution does
it move toward the smaller values which apply at horizon crossing.
Since $\ln k \sim N$ during inflation, the power grows in the past
like $P\sim e^{(n_s-1)N}$ to some approximation, where $n_s$ is
evaluated at the saddle point.  Knowing that $P(k_0) = 4\times
10^{-10}$ at the COBE point, we can thus roughly estimate the
number of $e$-foldings $N_b$ from the end of the stage of eternal
inflation, where $P = O(1)$, to the COBE point, where
$P(k_0) \sim  4\times 10^{-10}$: 
\beq
\label{approx_efolds}
    N_b \sim {\ln(4\times 10^{-10})\over n_s-1} = {C\over 1-n_s}\ ,
\eeq
with $C \approx 20$. In the model  with $W_0 = 5.227\times 10^{-6}$
one has $1-n_s \approx 0.022$ during the main part of inflation (much
earlier than horizon crossing), which gives $N_b \approx 980$, in
reasonable agreement with the actual value of 920 obtained 
numerically.  (Note, however, that even though $n_s = 0.98$ near the
saddle point, its value at the COBE scale is $n_s = 0.95$.). On the 
other hand, eq. (\ref{approx_efolds}) suggests that there will be a 
smaller number of $e$-foldings as we decrease the value of $n_s$. In fact,
we have proved explicitly that this is indeed what happens in our model. 
Solving the equations of motion numerically we found that the number of 
$e$-folds after eternal inflation for the case with $n_s=0.92$ gives us $300$
$e$-folds instead of the $920$ of the $n_s=0.95$ case.

This means that the total volume of the universe after eternal
inflation grows 
additionally by the factor of
\beq
    e^{3N_b} \sim \exp \left({3C\over 1-n_s}\right) .
\eeq
One could arrive at a similar conclusion by investigating the total
growth of the universe for inflation which begins when the field was
at a given distance from the saddle point. 

The meaning of this result is very simple: The small value of 
$1-n_{s}$ correspond to small values of the slow-roll parameters. 
One may argue that in this case eternal inflation near the saddle 
point becomes more efficient (``longer eternity''). In addition 
to it, the normal inflationary regime becomes longer, the growth 
of the volume of the universe is proportional to 
$ \exp \left({3C\over 1-n_s}\right)$, and therefore the total 
volume of the universe will be dominated by the regions with 
the smallest possible values of $1-n_{s}$. 

This argument  has a following interesting implication, 
which seems quite plausible: If one assumes
that the probability to live in a given part of the universe is
proportional to its volume,  this assumption singles out those parameters which
lead to $n_{s}$ as close to 1 as possible. In this sense, one may
argue that the value $W_0 = 5.227\times 10^{-6}$, which gives the
largest value of $n_{s}$ in our model (all other parameters being
fixed), is determined not by fine-tuning but by anthropic
considerations: The parts of the universe with a flat spectrum of
perturbations tend to have larger volume; the flatter the better.

\section{Discussion}\label{discussion}

We have seen that the superpotential and K\"ahler potential which
have been computed for compactifications to four dimensions on the
orbifolded Calabi-Yau manifold, $\IP^4_{[1,1,1,6,9]}$, is rich
enough to allow inflationary regimes to occur. These inflationary
regimes resemble those of the original racetrack scenario inasmuch
as they correspond to regions where $|\eta|$ is arranged to be
small at a saddle point which lies between two degenerate minima.

In the KKLT model, with the superpotential containing only one
exponent for the volume modulus, one could not have inflation without
adding moving branes. In our original racetrack inflation scenario
\cite{racetrack} we were able to find the first working inflationary
model without adding any new branes to the KKLT vacuum stabilization
scenario. In the present work we have made a new step and   achieved
inflation in a theory with two moduli fields, without introducing
the standard racetrack potentials with two exponential terms for each
of them. This suggests that by increasing the number of
moduli fields, inflation may be  easier to achieve.  Actually, in
Ref. \cite{conlon} inflation is generically achieved in models with more than two
K\"ahler moduli, but  the inflaton in those models is
 the real part of a K\"ahler
modulus rather than its axionic component. A
similar observation was made in Ref. \cite{Dimopoulos:2005ac} in a
different context.

The properties of the resulting density fluctuations in our model can
be computed in the usual way in terms of the slow-roll parameters,
leading to inflationary phenomenology which resembles in many ways
that of the original racetrack inflation model. Depending on the
choice of parameters for the improved racetrack, one can obtain
spectra with different values of the spectral index $n_{s}$. We find
it interesting that the volume of the universe  becomes exponentially
larger for the models with the smallest deviation of $n_{s}$ from
unity. For the full range of inflationary parameters that  we were
able to find, the largest value of $n_{s}$, corresponding to the
largest volume of the universe, is $n_{s} \approx 0.95$, which is in
a good agreement with the latest observational data
\cite{MacTavish:2005yk}.

Finally, we would like to comment on the non-Gaussianity of the metric 
perturbations in our model. As should be clear from the discussion above, 
the model we are presenting can be recast as a single field inflation, 
once the trajectory in field space is obtained numerically. Based on this
realization one would expect the non-Gaussianity in our model to
be very small, and we checked numerically that this is indeed the
case.\footnote{Since completing this paper, there appeared
a claim \cite{Sun-Zhang} that our model is ruled out by the
generation of large non-Gaussian perturbations. We have discussed
this with the authors, and they now agree that the degree of nongaussianity
in our model is very small, in agreement with observations.}

\section*{Acknowledgments}
Seven of the Racetrack 8 would like to thank the Aspen Center for
Physics for their hospitality, when some of the early parts of
this work was done. We acknowledge most useful discussions with 
A.\ Iglesias, J.\ Conlon, F.\ Denef, B.\ Florea, C.\ Gordon, S.\
Kachru and M.\ Redi. J.J. B-P. is supported by the James Arthur 
Fellowship at NYU. C.B.'s research is supported in part by a grant 
from N.S.E.R.C. of Canada, as well as funds from McMaster University, 
the Perimeter Institute and the Killam Foundation. J.C.\ is supported 
by NSERC and FQRNT (Qu\'ebec). The work of R.K. and A.L. was supported 
by NSF grant PHY-0244728. The work of M.G.-R. was supported by the 
European Commission under contract MRTN-CT-2004-005104. F.Q. is
partially funded by PPARC and a Royal Society Wolfson award.

\section*{Note Added} This paper was completed shortly before the
WMAP third year data release, and our results (including the
prediction $n_s=0.95$) were reported by A.L.\ at the 
UK Cosmology Meeting, Newcastle University, March 14 2006.

\end{document}